\documentclass[aps,prd,twocolumn,floats,showpacs, nofootinbib]{revtex4}
\usepackage{graphics,epsfig,color} 
\usepackage{amsmath,amssymb}
\usepackage{hyperref}

\def\d {\mbox{d}}
\def\L {\mathcal{L}}

\def\be{\begin{equation}}
\def\ee{\end{equation}}
\def\bea{\begin{eqnarray}}
\def\eea{\end{eqnarray}}
\begin{document}
\title{Conservation laws in theories with universal gravity/matter coupling}
\author{Olivier Minazzoli}
\affiliation{UMR ARTEMIS, CNRS, University of Nice Sophia-Antipolis,
Observatoire de la C\^ote d’Azur, BP4229, 06304, Nice Cedex 4, France}
\affiliation{Jet Propulsion Laboratory, California Institute of Technology,\\
4800 Oak Grove Drive, Pasadena, CA 91109-0899, USA}
\begin{abstract}
In this paper, we investigate the conservation laws of different type of particles in theories with a universal gravity/matter coupling. The result brings new insights about previous studies on universal gravity/matter theories. Especially, the paper demonstrates that for perfect fluids, there is an equivalence between the assumption $\L_m=-\epsilon$, where $\epsilon$ is the total energy density; and the assumption that the matter fluid current is conserved ($\nabla_\sigma (\rho u^\sigma)=0$, where $\rho$ is the rest mass density). However, the main result is given in the general case where one does not make any assumption on the conservation of the matter fluid current. 
\end{abstract}
\pacs{04.50.Kd, 04.40.Nr, 04.20.Fy, 95.36.+x}
\maketitle


\section{Introduction}

Recently, many phenomenological theories of gravitation have been proposed with a universal gravity/matter coupling. The coupling appears either as a universal non-minimal coupling between the Ricci scalar curvature and the non-gravitational fields  \cite{bertolamiPRD07,bertolamiPRD08,sotiriouCQG08,bisabrPRD12b}; or as a universal non-minimal coupling between a scalar-field and the non-gravitational fields \cite{dasPRD08,avilesPRD11,moiARXIV12,atkinsPRD13}. In order to study both types of theories, we define universal gravity/matter coupling theories by the following action:
\begin{equation}\label{eq:action}
S=\frac{1}{c} \int \left[ f\left(R, \phi ,\left(\nabla \phi \right)^2 \right) + h\left(R, \phi ,\left(\nabla \phi \right)^2 \right) \L_m \right] \sqrt{-g} d^{4}x,
\end{equation}
where $g$ is the determinant of the metric tensor $g_{\mu \nu}$, $f(R,\phi,\left(\nabla \phi \right)^2)$ and $h(R,\phi,\left(\nabla \phi \right)^2)$ are arbitrary analytical functions of the Ricci scalar $R$, a scalar field $\phi $, and the gradients constructed from the scalar field and $\L_m$ is the Lagrangian of the non-gravitational sector -- that we assume here to be the usual standard model of particle Lagrangian\footnote{This assumption can be relaxed to include potential dark matter fields (see discussion in the conclusion).}. We dub the coupling \textit{universal} because the coupling is the same for the whole material sector; while a more general set of theories would require different coupling for each material field \cite{armendarizPRD02,DamPolyNPB94,olivePRD08}. It is worth stressing that such an action encompasses Brans-Dicke-like scalar-tensor theories (with \cite{dasPRD08,avilesPRD11,moiARXIV12,atkinsPRD13} and without \cite{bransAIPC08,PeriPRD10,deviPRD11,saez-gomezPRD12} universal scalar/matter coupling), $f(R)$ theories (with \cite{bertolamiPRD07,bertolamiPRD08,sotiriouCQG08} and without \cite{sotiriouRVMP10} universal $R$/matter coupling), some low energy (tree level) string inspired 4-dimensional models in the string frame \cite{DamPolyGRG94,damourPRD96,gasperiniPRD12}, several Kaluza-Klein theories reduced to 4 dimensions \cite{coquereauxCQG90,overduinPhR97,senseiPRD07} and most of the models present in $ f\left(R,\L_m, \phi ,\left(\nabla \phi \right)^2 \right)$ theories \cite{harkoPRD13}. However, we consider only theories where $\phi$ is a gravitational field -- which requires some sort of coupling between $\phi$ and R, as in Brans-Dicke-like scalar-tensor theories for instance \cite{bransPhRv61,bransPhRv62}. Moreover, the main feature of the action we are interested in is due to the non-constant term $h(R,\phi,\left(\nabla \phi \right)^2)$. This last fact explains the chosen name for this kind of non-minimal coupling: \textit{gravity/matter coupling}. 

This type of universal gravity/matter coupling has been proposed in order to explain the phenomena usually associated to \textit{dark matter} (without modifications of the standard model of particles) \cite{moffatJCAP06,bertolamiPRD07,moffatCQG09,harkoPRD10b,bertolamiPRD12,moffatIJMPD12}, to mimic the cosmological constant \cite{bertolamiPRD11}, to give a reheating scenario in the context of Starobinsky inflation \cite{bertolamiPRD11b}, to explain the acceleration of the expansion of the Universe \cite{dasPRD08,bertolamiPRD10,gasperiniPRD12,bisabrPRD12, bisabrPRD12b,harkoIJMPD12}, to predict (essentially) stable WIMPs \cite{damourPRD96}, or to explain the smallness of the gravitational constant $G$ compared to the proton mass scale \cite{cliftonPRD06}. Moreover, it has been shown that under specific assumptions such theories can be compatible with current tight constraints on the equivalence principle(s) (see for instance \cite{DamPolyNPB94,damourPRL02,damourPRD10,damourCQG12}). In the end, the goal of such a phenomenological action is to tend to an explanation of the whole observed phenomena, without the need of separate theories for inflation, dark matter or dark energy. 

As first noticed in \cite{bertolamiPRD07,sotiriouCQG08,bertolamiPRD08} in the case of $f(R)$ with universal gravity/matter coupling theories, the field equations of theories with a gravity/matter coupling depend explicitly on the \textit{on-shell} Lagrangian $\L_m$. In particular, whether one chooses $\L_m=\epsilon$  -- where $\epsilon$ is the total energy density \cite{moi_hPRD12} -- or $\L_m=P$ -- where $P$ is the pressure of the perfect fluid --  (or any normalized linear combination of the two) explicitly modifies the field equation; while the difference between the different \textit{on-shell} Lagrangians should reduce to surface integrals that do not contribute to the field equations -- just as in the case of general relativity where one can freely choose between $\L_m=\epsilon$ and $\L_m=P$ without modifying the field equations \cite{brownCQG93}. Therefore it was first thought that there was a degeneracy of supposedly equivalent actions that lead to different field equations \cite{bertolamiPRD08} -- which is not satisfactory with respect to the monist view of modern Physics, which requires a unique mathematical description of the natural phenomena. However, \cite{moi_hPRD12} showed that some assumptions on the nature of the matter fluid can kill the degeneracy, hence showing that different Lagrangians actually correspond to matter fluids with different fundamental characteristics. To be specific, \cite{moi_hPRD12} showed that when the matter fluid current is conserved, the Lagrangian of a barotropic perfect fluid  is $\mathcal{L}_m=-\epsilon$ -- and not anything else. In the following we discuss this issue in more details and generalize the result found in \cite{moi_hPRD12} to general perfect fluids where the matter fluid current is not conserved.

In section \ref{sec:fieldEQ}, we give the field equations corresponding to the action (\ref{eq:action}). After, in section \ref{sec:PF} we study the equation of motion of interacting particles forming an arbitrary perfect fluid. Then, in section \ref{sec:ppmonop} we derive the equation of motion of non-interacting point-particles. Finally, in section \ref{sec:EW} we derive the equation of motion of light rays in the geometric optic approximation before concluding in section \ref{sec:concl}.

\section{Field equations of the universal gravity/matter coupling model}
\label{sec:fieldEQ}

We use the usual definition of the stress-energy tensor:
\begin{equation}
T _{\mu \nu }=-\frac{2}{\sqrt{-g}}\frac{\delta \left( \sqrt{-g} \L_{m}\right)}{
\delta g^{\mu \nu }}\,.
\end{equation}
By assuming that the matter Lagrangian $\L_{m}$ depends only
on the metric tensor components $g_{\mu \nu }$, and not on its derivatives,
we obtain the stress-energy tensor as
\begin{equation}\label{en1}
T_{\mu \nu }=g_{\mu \nu }\L_{m}-2\frac{\delta \L_{m}}{\delta g^{\mu
\nu }}.
\end{equation}
The extremization of the action (\ref{eq:action}) then reduces to the following field equations \cite{harkoPRD13}:
\begin{eqnarray}\label{eq:field}
 \left[f_R+h_R \L_m\right] R_{\mu \nu }+\left( g_{\mu \nu }\nabla _{\lambda }\nabla^{\lambda }
-\nabla
_{\mu }\nabla _{\nu }\right) \left[f_R+h_R \L_m\right]
    \nonumber  \\
- \frac{1}{2}f g_{\mu \nu }=\frac{1}{2}%
h~ T_{\mu \nu }-\left( f_{\left(\nabla \phi \right)^2 }+h_{\left(\nabla \phi \right)^2 } \L_m \right)\nabla _{\mu }\phi \nabla _{\nu }\phi ,\nonumber\\
\end{eqnarray}
where the subscript of $f$ or $h$ denotes a partial derivative with respect to the arguments, i.e., $f_R=\partial f/\partial R$, $f_{\left(\nabla \phi \right)^2 }=\partial f/\partial \left(\nabla \phi \right)^2$.
\begin{equation}
\square _{\left( \nabla \phi \right) ^{2}}\phi =\frac{1}{2} \left( f_{\phi }+h_{\phi} \L_m \right),
\end{equation}
where $f_\phi=\partial f/\partial \phi$ and 
\begin{equation}\label{2}
\square _{\left( \nabla \phi \right) ^{2}}=\frac{1}{\sqrt{-g}}\frac{\partial
}{\partial x^{\mu }}\left[\left( f_{\left(\nabla \phi \right)^2 }+h_{\left(\nabla \phi \right)^2 } \L_m \right)\sqrt{-g}%
g^{\mu \nu }\frac{\partial }{\partial x^{\nu }}\right] .
\end{equation}
The (non-)conservation of the stress-energy tensor therefore reduces to:
\be
\nabla_\sigma T^{\mu \sigma} = \left(\L_m g^{\mu \sigma} - T^{\mu \sigma} \right) \partial_\sigma \ln{h} \label{eq:non_conserv}.
\ee

\section{Perfect fluid particles}
\label{sec:PF}

The general Lagrangian that leads to the stress-energy tensor of a perfect fluid is 
\be
\L_m=[-(1-\alpha) \epsilon + \beta P] \gamma^{-1}, \label{eq:PFLangran}
\ee
where $\alpha$, $\beta$ are constants, and $\gamma$ is a normalization constant such that $\gamma=1-\alpha+\beta$, and where $P$ and $\epsilon$ are the pressure and the so-called total energy density of the fluid \cite{fockBOOK64}. Indeed, following the development made in \cite{brownCQG93,bertolamiPRD08}, one can show that $\L_m=[-(1-\alpha) \epsilon + \beta P] \gamma^{-1}$, in addition with (\ref{en1}) induces $T^{\alpha \beta}=(\epsilon+P) U^\alpha U^\beta+P g^{\alpha \beta}$ -- see appendix of \cite{moi_hPRD12} for an example. The normalization procedure is necessary in order to conserve energy through change of parametrization. Indeed, adding surface integral terms, one can transform the \textit{on-shell} Lagrangian from $-\epsilon$ to $P$ (vice versa). Therefore, the most general Lagrangian has to be a normalized sum of the two possible \textit{on-shell} Lagrangians. For instance, for $P=0$, the normalization induces that $\L_m=-c^2 \rho$ for any $\alpha$; while $\L_m$ could be any times the rest mass density if not normalized. On the other side, one should note that the normalization procedure discards $\L_m=T$ as a possible \textit{on-shell} Lagrangian for fluids with $P \neq 0$. The corresponding normalized \textit{on-shell} Lagrangian would write $\L_m=T/4$ instead. However, from the trace of equation (\ref{en1}), one gets that $\L_m=T/4$ implies that $\delta \L_{m}/\delta g^{\mu \nu }$ has to be trace-free. Therefore, $\L_m=T/4$ is not suitable any kind of effective particles.

Also, as shown in appendix of \cite{moi_hPRD12}, the matter current is generally not conserved when considering $\L_m=P$. Therefore we deduce that the matter current is generally not conserved when considering the Lagrangian (\ref{eq:PFLangran}). Thus, let us write $\nabla_\sigma (\rho U^\sigma)=D$, where $D$ is any scalar to be determined. One deduces
\be
\nabla_\sigma (\epsilon U^\sigma)=\left( \frac{\epsilon+P}{\rho} \right) D - P \nabla_\sigma U^\sigma,
\ee
where  \cite{moi_hPRD12}
\be
\epsilon= \rho \left(c^2 - \frac{P}{\rho}+ \int \frac{\d P}{\rho} \right). 
\ee
Therefore, taking the divergence of $T^{\alpha \beta}=(\epsilon+P) U^\alpha U^\beta+P g^{\alpha \beta}$, one gets:
\bea
\nabla_\sigma T^{\mu \sigma}&=& \left(\epsilon+P \right) U^\sigma \nabla_\sigma U^\mu + \left(g^{\mu \sigma}+U^{\mu} U^{\sigma} \right) \nabla_\sigma P \nonumber\\
 &&+ U^\mu \left(\frac{\epsilon+P}{\rho} \right)  D. \label{eq:grad1}
\eea
On the other side, equation (\ref{eq:non_conserv}) implies:
\bea
\nabla_\sigma T^{\mu \sigma}&=& -\left(\epsilon+P \right) \left[g^{\mu \sigma}+U^\mu U^\sigma \right] \partial_\sigma \ln h  \nonumber\\
 &&+ \left[\left(1- \frac{1-\alpha}{\gamma} \right) \epsilon + \frac{\beta}{\gamma} P \right]g^{\mu \sigma} \partial_\sigma \ln h.\label{eq:grad2}
\eea
Multiplying equations (\ref{eq:grad1}) and  (\ref{eq:grad2}) by $U_\mu$ and equating them, one gets:
\be
\nabla_\sigma (\rho U^\sigma) = - \frac{\rho}{\epsilon+P} \left[\left( 1 - \frac{1-\alpha}{\gamma} \right) \epsilon + \frac{\beta}{\gamma} P \right] U^\sigma \partial_\sigma \ln h. \label{eq:nonconsab}
\ee

One note that $\alpha=\beta=0$ induces $\nabla_\sigma (\rho u^\sigma)=0$ -- independently of the value of $\partial_\alpha \ln h$. Thus, it proves that $\L_m =-\epsilon \Rightarrow \nabla_\sigma(\rho u^\sigma)=0$ for all perfect fluids. Since \cite{moi_hPRD12} proves $\nabla_\sigma(\rho u^\sigma)=0 \Rightarrow \L_m =-\epsilon$ for barotropic perfect fluids, one has $\L_m =-\epsilon \Leftrightarrow \nabla_\sigma(\rho u^\sigma)=0$ (at least) for barotropic perfect fluids. Now, injecting (\ref{eq:nonconsab}) in (\ref{eq:non_conserv}) gives:
\begin{widetext}
\begin{center}
\be
\left[\epsilon + P \right] U^\sigma \nabla_\sigma U^\alpha = \left(g^{\alpha \sigma}+U^\alpha U^\sigma \right) \left[\partial_\sigma P + \left(-\frac{1-\alpha}{\gamma} \epsilon + \frac{\beta}{\gamma} P \right) \partial_\sigma \ln h \right].\label{eq:devGEN}
\ee
\end{center}
\end{widetext}
This equation generalizes the result found in \cite{roshanPRD13} for non-interacting monopoles to interacting particles in a perfect fluid. 

\section{Monopole test particles}
\label{sec:ppmonop}
In this section, we derive the equation of motion of point particles directly from the material sector of the action -- while \cite{roshanPRD13} starts from the (non-)conservation equation of the stress-energy tensor. Eventually, we shall show the equivalence of the two approaches. The material part of the action writes:
\begin{equation}\label{eq:actionm}
S_m=\frac{1}{c}\int h(R, \phi ,\left(\nabla \phi \right)^2)~ \L_m  \sqrt{-g} d^{4}x.
\end{equation}
For a non-interacting point particle one has $\L_m = -c^2 \rho$, with $\rho = \mu \delta(\vec{x})$ ($\mu$ being a mass), such that
\be
S_m=- c \int_W  \sqrt{-g}\mu~ h~ cdt=- c \int_W  \sqrt{-g}\mu u^0~ h~ ds, \label{eq:red_action}
\ee
where the integral is taken on the world line W of the particle, and where $u^\alpha=dx^\alpha/ds$, $s$ being an affine parameter of the world line. According the to the last section, since non-interacting point-particles are the simplest form of perfect fluids ($P=0$), the Lagrangian $\L_m = -c^2 \rho$ induces the conservation of the matter fluid current $\nabla_\sigma \left(\rho u^\sigma \right)=0$. Therefore, one shows that $\L_m=-c^2 \rho$ induces the Newtonian conservation of the so-called \textit{conserved density} $\rho^*= \sqrt{-g} \rho u^0$ \cite{brumbergNCimB89}:
\be
\partial_0 \left(\rho^* \right)+\partial_i (\rho^* v^i)=0,
\ee
where $v^i=u^i/u^0$. Hence, one deduces the conservation of the mass $m$ ($dm/dx^0 =(u^0)^{-1} dm/ds=0$) defined as $m=\sqrt{-g} \mu u^0$. One has
\be
\nabla_\sigma \left(\rho u^\sigma \right)=0 \Leftrightarrow \frac{\d m}{\d s}=0.
\ee
Therefore, one can take $m$ out of the integral in (\ref{eq:red_action}) and one gets the modified point particle action:
\be
S_m=-m c \int_W  h~ ds.
\ee
Parametrizing by the proper time $\tau$, one gets $S_m=-m c^2 \int_W  L d\tau$, with $L=h \sqrt{g_{\alpha \beta} U^\alpha U^\beta}$, with $U^\alpha =c^{-1} dx^\alpha/d\tau$; such that one can use the Euler-Lagrange equation:
\be
\frac{\partial L}{\partial x^\gamma}-\frac{\d}{c\d \tau} \frac{\partial L}{\partial U^\gamma}=0.
\ee
Computing the following terms,
\be
\frac{\partial L}{\partial x^\gamma}=\frac{\d s}{c\d \tau} \partial_\gamma h + \frac{1}{2} \frac{c\d \tau}{\d s} h \partial_\gamma g_{\alpha \beta} U^\alpha U^\beta,
\ee
and
\bea
\frac{\d}{c\d \tau}&& \frac{\partial L}{\partial U^\gamma}=  \frac{c\d \tau}{\d s} \times \\ 
&&\left(g_{\alpha \gamma} \partial_\beta h ~U^\alpha U^\beta + h \partial_\beta g_{\alpha \gamma}~ U^\alpha U^{\beta} + h g_{\alpha \gamma} \frac{\d U^\alpha}{c\d \tau} \right), \nonumber
\eea
one gets the modified equation of motion of a point particle:
\be
\frac{\mbox{D} U^\alpha}{c\d \tau}=- \left(g^{\alpha \sigma}+U^\alpha U^\sigma \right) \partial_\sigma \ln h, \label{eq:modmonop}
\ee
where $\mbox{D}/\d \tau$ stands for the usual covariant derivative ($\mbox{D}/\d \tau \equiv c U^\sigma \nabla_\sigma$). One can check that this equation is in accordance with the result found in \cite{roshanPRD13}, and that for $P=0$, equation (\ref{eq:devGEN}) reduces to (\ref{eq:modmonop}). It shows that, as expected, deriving the particle equation of motion directly from the Lagrangian is equivalent to deriving them from the (non-)conservation equation of the stress-energy tensor.

One should note that thanks to the conservation of the matter fluid current, one can unambiguously define an invariant mass $m$ on the particle's worldline. Then, one can define an \textit{apparent inertial mass} $M$ as:
\be
M\left(R, \phi ,\left(\nabla \phi \right)^2 \right)=m~ h\left(R, \phi ,\left(\nabla \phi \right)^2 \right),
\ee
such that the material action takes the usual form:
\be
S_m=- c \int_W  M~ ds. \label{eq:actionMinert}
\ee
This is of course in disagreement with the Einstein equivalence principle. However, one can note that the mass defined in (\ref{eq:actionMinert}) is constant in non-gravitational regimes (ie. the mass $M$ of particles is constant at small enough scales where gravitation can be neglected -- and when the experiments are made in a small enough period of time such that the cosmologically driven possible variations of the background value of the gravitational fields $g_{\mu \nu}$ and $\Phi$ are negligible \cite{DamPolyGRG94}). In particular, one should note that specific cases can satisfy the present tight constraints on the equivalence principle(s) (see for instance \cite{DamPolyNPB94,damourPRL02,damourPRD10,damourCQG12}).
\section{Electromagnetic waves in the geometric optic regime}
\label{sec:EW}

In the following, the electromagnetic field is not considered as a significant source of curvature (ie. photons are considered as test particles). Then, from (\ref{eq:action}) the electromagnetic equation writes:
\begin{equation}
\nabla_\sigma \left(h(R,\phi,\left(\nabla \phi \right)^2) ~F^{\mu \sigma} \right)=0.
\end{equation}

Using the Lorenz Gauge ($\nabla_\sigma A^\sigma =0$), it reduces to:
\begin{equation}
-\Box A^\mu + g^{\mu \epsilon} R_{\gamma \epsilon} A^\gamma + \left(\nabla^\mu A^\sigma - \nabla^\sigma A^\mu \right) \partial_\sigma \ln h=0. \label{eq:maxmodLG}
\end{equation}

Following the analysis made in \cite{MTW}, we expand the 4-vector potential as follows:
\begin{equation}
A^\mu = \Re \left\{ \left(a^\mu + \epsilon b^\mu + O(\epsilon^2) \right) \exp^{i \theta / \epsilon} \right\},
\end{equation}
The two first leading orders of equation (\ref{eq:maxmodLG}) respectively give:
\begin{equation}
k_\sigma k^\sigma =0, 
\end{equation}
where $k_\sigma \equiv \partial_\sigma \theta$, and
\begin{equation}
a^\mu \nabla_\sigma k^\sigma + 2 k^\sigma \nabla_\sigma a^\mu =  \left(k^\mu a^\sigma - k^\sigma a^\mu \right) \partial_\sigma \ln h.
\end{equation}
Remembering that the Lorenz Gauge condition gives $k_\sigma a^\sigma=0$ at the leading order, one gets:
\begin{equation}
k^\sigma \nabla_\sigma k^\mu = 0.
\end{equation}
This equation is the usual null-geodesic equation, showing that the presence of non-minimal gravity/matter coupling won't affect light ray trajectories at the geometric optic approximation. However, defining $a^\mu = a f^\mu$, the propagation equation for the scalar amplitude ($a$) as well as the propagation equation for the polarization vector ($f^\mu$) are modified:
\begin{eqnarray}
&&k^\sigma \nabla_\sigma a = - \frac{a}{2} \nabla_\sigma k^\sigma + \frac{1}{2} a k^\sigma \partial_\sigma \ln h, \\
&&k^\sigma \nabla_\sigma f^\mu = \frac{1}{2} k^\mu f^\sigma \partial_\sigma \ln h.
\end{eqnarray}
From there follows that the conservation law of te "photon number" (ie. intensity) is modified:
\begin{equation}
\nabla_\sigma (k^\sigma a^2)= - a^2 k^\sigma \partial_\sigma \ln h. \label{eq:NCintensity}
\ee
One notes that the last three equations may give alternative ways to put constraints on those theories.

For instance, at the classical level, equation (\ref{eq:NCintensity}) leads to a modification of the distance luminosity $\d_L$ because of the energy transfer between the gravitational fields and the electromagnetic field. As an example, for a flat FRLW metric, the new equation reads \footnote{This equation has been obtained in a parallel work, in collaboration with Aurelien Hees (unpublished at the present time).}:
\be
\d_{L,k=0} = (1+z)\sqrt{\frac{h|_{z=0}}{h|_z}} \int_0^z \frac{dz}{H(z)} , \label{eq:diffDLdeff}
\ee 
where $H(z)$ is the Hubble parameter as it would be measured by an observer at redshift $z$. This has an implications in observational cosmology since the accelerated expansion of the Universe is historically deduced from the observation of high redshift supernovae  while assuming the usual equation for the distance luminosity \cite{perlmutterApJ99}:
\be
\d_{L,k=0 \mathrm{-usual}} = (1+z) \int_0^z \frac{dz}{H(z)}.
\ee  
Therefore in models represented by the action (\ref{eq:action}), one not only has to check whether or not the Universe can be accelerated (such as in \cite{dasPRD08,bertolamiPRD10,bisabrPRD12b,bisabrPRD12}); but one also has to pay attention on how the distance luminosity equation is affected by the evolution of the Universe when considering (\ref{eq:diffDLdeff}). As far as we know, no study took that point into account so far.

At the quantum level, equation (\ref{eq:NCintensity}) suggests that an excess/shortage of photons with momentum parallel to the gradient of the gravitational potential should be expected at some level. However, this question requires a dedicated QED study in each specific theory that is included in the general action (\ref{eq:action}), in order to see whether or not this effect could be significant in some current experiments or observations.

\section{Conclusion}
\label{sec:concl}

In the present paper we studied the motion of non-interacting point-particles, interacting particles composing a general perfect fluid and the motion of light rays in theories that exhibit a universal gravity/matter coupling. We saw that the transfer of energy between the gravity fields and the material fields -- that is due the universal gravity/matter coupling -- depends on the nature of the material fields. Indeed such a transfer can take two possible forms: either it modifies the matter fluid current conservation, or it modifies the equation of motion (or both at the same time). As an example, it seems that for massive non-interacting point-particles, the energy transfer is totally incorporated into the modification of the equation of motion; while for light rays, the energy transfer is totally incorporated into the modification of the conservation of the intensity. Otherwise, we saw that for a very general set of perfect fluids, the energy transfer modifies both the matter current fluid conservation and the equation of motion. However, all the perfect fluid Lagrangians that give a non-conservation of the rest mass seem to be in physical disagreement with the non-interacting point-particle case. Indeed, using the relation $\L_m=-\epsilon \Rightarrow \nabla_\sigma(\rho U^\sigma)=0$  demonstrated for perfect fluids in section \ref{sec:PF}, the non-interacting point-particle case necessarily induces the conservation of the rest mass (see section \ref{sec:ppmonop}). This seems to be in accordance with the assumption that $\L_m$ is the usual material Lagrangian. Therefore, we argue that perfect fluid cases where $\L_m=[-(1-\alpha) \epsilon + \beta P] \gamma^{-1}$, with $\alpha \neq 0$ and $\beta \neq 0$, might not have any physical significance -- unless one modifies the material sector in such a specific way that the non-conservation of matter precisely reduces to (\ref{eq:nonconsab}), with $\alpha \neq 0$ and $\beta \neq 0$. Hence, since the last case seems very unlikely (unless maybe for some exotic material fields), we believe that theorists should be careful when using Lagrangians that are such that $\L_m=[-(1-\alpha) \epsilon + \beta P] \gamma^{-1}$, with $\alpha \neq 0$ and $\beta \neq 0$. This point is very important if one considers the numerous publications using various values for the parameters $\alpha$ and $\beta$ without justifications  (see in appendix \ref{sec:sample} a sample of cases found in the literature). However, this issue would get a definitive answer if, instead of using a phenomenological perfect fluid description of matter, one works with the actual fundamental fields of the material sector (for instance, with $\L_m$ describing the standard model of particle fields). Work in that direction is in progress.

Also, although (\ref{eq:action}) generally leads to a violation of the equivalence principle(s), specific cases can still be in accordance with the present tight experimental constraints (see for instance \cite{DamPolyNPB94,damourPRL02,damourPRD10,damourCQG12}). However a dedicated study of the equivalence principle issue in each specific case is mandatory.

\appendix

\section{Perfect fluid models in the literature}
\label{sec:sample}

\begin{center}
\begin{tabular}{|l |l|l| }
\hline
\multicolumn{3}{ |c| }{Sample of cases considered in the literature} \\
\hline
Bertolami et al. \cite{bertolamiPRD12} & $\alpha=0, \beta=0$, $\gamma=1$  & $dm/ds = 0$ \\
\hline
Minazzoli and Harko \cite{moi_hPRD12} & $\alpha=0, \beta=0$, $\gamma=1$  & $dm/ds=0$\\
\hline
Harko et al. \cite{harkoPRD13} & $\alpha=0, \beta=0$, $\gamma=1$  & $dm/ds=0$ \\
\hline
Sotiriou and Faraoni \cite{sotiriouCQG08} & $\alpha=1, \beta=1$, $\gamma=1$  & $dm/ds\neq 0$ \\
\hline
Bisabr \cite{bisabrPRD12,bisabrPRD12b} & $\alpha=1, \beta=1$, $\gamma=1$  & $dm/ds\neq 0$ \\
\hline
Farajollahi and Salehi \cite{farajollahiJCAP10} & $\alpha=0, \beta=3$, $\gamma=4$  & $dm/ds\neq 0$ \\
\hline
Jamil et al. \cite{jamilEPJP11} & $\alpha=0, \beta=3$, $\gamma=4$  & $dm/ds\neq 0$ \\
\hline
Sheykhi and Jamil \cite{sheykhiPLB11} & $\alpha=0, \beta=3$, $\gamma=4$   & $dm/ds\neq 0$ \\
\hline
Saaidi et al. \cite{saaidiPRD11} & $\alpha=0, \beta=3$, $\gamma=4$  & $dm/ds\neq 0$ \\
\hline
Sharif and Waheed \cite{sharifIJMPD12} & $\alpha=0, \beta=3$, $\gamma=4$ & $dm/ds\neq 0$ \\
\hline
\end{tabular}
\end{center}

\begin{acknowledgments}
This research was partly done as an invited researcher of the Observatoire de la C\^ote d'Azur.\\
This research was partly supported by an appointment to the NASA Postdoctoral Program at the Jet Propulsion Laboratory, California Institute of Technology, administered by Oak Ridge Associated Universities through a contract with NASA. Government sponsorship acknowledged. \\
The author wants to thank Aurelien Hees, Tiberiu Harko, Francisco Lobo and Viktor Toth for reading the paper and for their interesting comments.

\end{acknowledgments}

\end{document}